\documentclass[prd,twocolumn]{revtex4}
\usepackage{graphicx, epsfig}
\usepackage{color}
\usepackage{mathrsfs}
\usepackage{amsmath}
\usepackage{latexsym}

\newcommand{\be}{\begin{equation}}
\newcommand{\ee}{\end{equation}}
\newcommand{\bea}{\begin{eqnarray}}
\newcommand{\eea}{\end{eqnarray}}

\newcommand{\gapp}{\mathrel{\raise.3ex\hbox{$>$}\mkern-14mu
              \lower0.6ex\hbox{$\sim$}}}
\newcommand{\lapp}{\mathrel{\raise.3ex\hbox{$<$}\mkern-14mu
              \lower0.6ex\hbox{$\sim$}}}

\begin{document}
\title{Hawking radiation of unparticles}
\author{De-Chang Dai and Dejan Stojkovic}
\affiliation{HEPCOS, Department of Physics,
SUNY at Buffalo, Buffalo, NY 14260-1500}


\begin{abstract}
\widetext
Unparticle degrees of freedom, no matter how weakly coupled to the standard model particles, must affect the evolution of a black hole, which  thermally decays into all available degrees of freedom. We develop a method for calculating the grey-body factors for scalar unparticles for $3+1$ and higher dimensional black holes. We find that the power emitted in unparticles may be quite different from the power emitted in ordinary particles. Depending on the parameters in the model, unparticles may become the dominant channel. This is of special interest for small primordial black holes and also in models with low scale quantum gravity where the experimental signature may significantly be affected. We also discuss the sensitivity of the results on the (currently unknown) unparticle normalization.
\end{abstract}


\maketitle

\section{Introduction}

There are many reasons to believe that interesting physics exists between the electroweak energy scale ($\sim$TeV) and the Planck energy scale ($\sim 10^{19}$GeV). Esthetic problem of the huge desert between the scales, the standard model hierarchy problem, unification of gauge couplings etc, all require new physics between these two scales. Many interesting models which accommodate new physics have been proposed. In the most recent development, the existence of a new scale invariant sector very weakly coupled to the standard model was postulated \cite{Georgi:2007ek,Georgi:2007si,Georgi:2008pq}. The fundamental energy scale, $M_F$, of this sector is perhaps far beyond the reach of today's or near future accelerators. However, the existence of such a sector may affect low energy phenomenology. The effective low energy field theory which describes these effects is often called unparticle physics since these new degrees of freedom would not behave as ordinary particles. For example, their scaling dimension does not have to be an integer or half an integer. Because of the scale invariance, the fundamental particles are massless.  An unparticle is a composite state of the fundamental massless particles, and couples to the standard model particles through a heavy mediator. This heavy mediator has a mass of the order of  $M_F$. Thus, interactions between the unparticles and the standard model particles is suppressed by powers of $M_F$. The energy scale at which unparticle physics is manifest is $\lambda_u$. At energies above $\lambda_u$, physics is more appropriately described in terms of fundamental particles rather than unparticles. The scale $\lambda_u$ of a "phase transition" in some sufficiently decoupled hidden sector could be high or low.  The presence of the standard model Higgs sector perhaps indicates that scale invariance is violated (at least for scalar unparticle degrees of freedom) below the electroweak scale and the unparticle behavior cannot arise below ${\cal O} (100)$ GeV.

Unparticles have many odd characteristics \cite{Georgi:2007si}. One of them is the continuous mass \cite{Nikolic:2008ax}. This fact  increases the effective number of species of unparticles. Although the interaction between unparticles and the standard model particles is supposed to be weak, the  number of species may overcome the suppression and the total effect can be large. A lot of work \cite{Zhang:2008zzy,He:2008xv,He:2008ef,kc,Kumar:2008zzb,Feng:2008ae,Kikuchi:2008pr,Kikuchi:2008nm,Kikuchi:2008fm,Boyanovsky:2008bf}
recently focused on the new collider signals for unparticle physics.
The unparticles could also play the role in the early universe  \cite{Davoudiasl:2007jr,Wei:2008nc,Lee:2009ny,Chen:2009ui,Chen:2009ys}, are a possible dark matter candidate \cite{Kikuchi:2007az,Kikuchi:2008zz} and can play the role of the quintessence dark energy field \cite{ude}. A tensor unparticle  mode can  modify Einstein's gravity \cite{Goldberg:2008zz}, in a way to increase the cross section of black hole production in high energy collisions \cite{Mureika:2007nc} or produce large density fluctuations. This in turn can increase the density of primordial black holes. For earlier work on black hole evaporation of composite degrees freedom see \cite{MacGibbon:1991tj}.

Finally, these new degrees of freedom can also change the power spectrum and lifetime of black holes. Black hole radiation is (approximately \cite{Vachaspati:2006ki}) thermal, which means that all available degrees of freedom are emitted. Suppressed interactions with the standard model particles play no role in this case, since gravity is blind to this suppression. Thus, black hole radiation will be modified, which in turn has the potential to put new constraints on the unparticle parameter space.

Here, we will focus on the Hawking radiation of the scalar unparticle. Because of the peculiar properties of unparticles (e.g. continuous mass), we have to modify the standard procedure for calculating gray-body factors. We will achieve this in two steps. In the first step, we will adapt the method of including the mass of the field in gray-body factor calculations.  Usually, for simplicity, the gray body factors are calculated for massless fields. This approximation is valid for small black holes whose temperature is much larger than the mass of the field, or in the last stages of evaporation of large black holes. For any other purpose, the mass of the field can not be neglected.
This is especially true in the case of unparticles where the effective unparticle mass changes continuously. In the second step, we will introduce the mass spectral density of unparticles and integrate over all of the possible degrees of freedom. The results indicate that the power spectrum strongly depends on the scaling dimension $d_u$ and unparticle energy scale. Since most of the unparticle degrees of freedom are light (with the mass concentrated around zero) a black hole of any size easily emits them. This is also true for any other massless scalar particle. The question is then in which regime the emitted unparticle spectrum is different from the standard scalar field spectrum. We find that power emitted in unparticles and ordinary particles is different in almost all frequency regimes. We note here that our calculations are valid only as long as one can treat unparticles as fundamental degrees of freedom. At the very last stages of evaporation, when the black hole temperature is higher than the unparticle phase transition scale $\lambda_u$, the effective unparticle description fails and we need a full fundamental theory.

\section{The model for Hawking radiation of unparticles}
\label{sec:model}
We consider a spherically symmetric black hole. The metric is written as
\begin{equation}\label{metric}
ds^2=-F(r)dt^2+\frac{1}{F(r)}dr^2+r^2d\Omega
\end{equation}
For a Schwarzschild black hole in a $3+1$ dimensional space-time we have $F(r)=1-r_g/r$, where the gravitational radius is $r_g=2GM/c^2$. The parameter M is the mass of the black hole, $G$ is Newton's gravitational constant and $c$ is speed of light.

We treat an unparticle as an effective massive scalar field. We will adapt the treatment of the massive scalar field from \cite{Firsova:1999jz,Goncharov:2000kg}. The wavefunction of a scalar field with the mass $\mu$ satisfies the Klein-Gordon equation
\be
\Box \Psi =\mu^2 \Psi
\ee
We assume that the variables can be separated as $\Psi=e^{-i\omega t}R(r) \Theta (\theta,\phi)$. The radial equation can then be written as
\begin{equation}
\label{master}
\frac{1}{r^2}\partial_r \left[Fr^2\partial_r R (r)\right]+\left[\frac{\omega^2}{F}-\frac{l(l+1)}{r^2}\right]R (r)=\mu^2 R(r)
\end{equation}
Here, $l(l+1)$ is the eigen value of the angular equation.
We follow the standard procedure for calculating gray-body factors and study the solutions to the radial equation in two limits, i.e.  near the gravitational radius and at infinity. These solutions are
\begin{eqnarray}
r\rightarrow r_g &:& R^{(r=r_g)} = A_{in}^{(r=r_g)}e^{-i\omega r_*} +A_{out}^{(r=r_g)}e^{i\omega r_*}\\
r\rightarrow \infty &:& R^{(r=\infty)} = A_{in}^{(r=\infty)}e^{-i\omega_1 r} +A_{out}^{(r=\infty)}e^{i\omega_1 r}
\end{eqnarray}
Here, $\omega_1=\sqrt{\omega^2-\mu^2}$. $r_*$ is a ``tortoise" coordinate defined by the solution to the equation $dr_*=dr/F$. We choose the boundary condition $A^{(r=r_g)}_{out}=0$ to ensure that near the horizon (i.e. gravitational radius) the solution is purely ingoing. We then numerically integrate Eq. (\ref{master}) using forth-order Runge-Kutta method. From the solution, we can calculate the absorption ratio as
\begin{equation}
|\tilde{A}_{l,m}|^2=1-\left|\frac{A^{(r=\infty)}_{out}}{A^{(r=\infty)}_{in}}\right|^2
\end{equation}
The power spectrum of the Hawking radiation is then written as
\begin{equation} \label{Emu}
\frac{d^2 E_\mu}{dtd\omega}=\sum_{l,m}\frac{\omega}{e^{\omega/T_h}-1}\frac{N_{l,m}|\tilde{A}_{l,m}|^2}{2\pi}
\end{equation}
The multiplicity of states is $N_{l,m}=2l+1$.

So far, we have not included the specific properties of unparticles.
We will do it what follows.

First we will define the mass spectral density (or mass distribution) $\rho(\mu)$ for unparticles. Since precise details of unparticle physics are not known, there will be ambiguity in normalization.

For an unparticle with the scaling dimension $d_u$ the mass spectral density with the general normalization is
\begin{equation} \label{rho}
\rho(\mu)=\frac{A_{d_u}\mu^{2(d_u-2)}}{\Lambda_u^{2(d_u-1)}}
\end{equation}
where the constant $\Lambda_u$ is inserted for dimensional reasons, and is in general arbitrary. Another arbitrary constant that depends only on $d_u$ is  $A_{d_u}$, which, depending on convention, may be absorbed into $\Lambda_u$.
We denote with $\lambda_u$ the scale of the phase transition below which unparticles appear. If we wish to normalize the unparticle degrees of freedom to unity, then we impose
\be
\int_0^{\lambda_u^2} \rho(\mu) d\mu^2 =1
\ee
which will be fulfilled for $\lambda_u = \Lambda_u$ and
\begin{equation} \label{rhou}
\rho(\mu)=\frac{(d_u-1)\mu^{2(d_u-2)}}{\Lambda_u^{2(d_u-1)}}
\end{equation}
The limit of $d_u \rightarrow 1$ is singular so $d_u$ must obey the strict inequality $d_u >1$.
The unit normalization is convenient because in the limit of $d_u \rightarrow 1$ the distribution in Eq.~(\ref{rhou}) becomes the Dirac delta function $\delta (\mu)$ and it reduces to the normalization of a single massless scalar field. However, this is very limiting since the exact unparticle combinatorics is currently unknown. Instead, the general normalization in Eq.~(\ref{rho}) (with $A_{d_u}$ absorbed into $\Lambda_u$) gives
\be \label{gn}
N=\int_0^{\lambda_u^2} \rho(\mu) d\mu^2 =\frac{1}{d_u-1}\left(\frac{\lambda_u}{\Lambda_u}\right)^{2(d_u-1)}
\ee
where $N$ is the total number of unparticle degrees of freedom. Clearly, $N$ can be very large in the limit of $d_u \rightarrow 1$. Since $N$ is currently unknown, we believe that the general normalization is more appropriate than the unit normalization. However, here we will compare results with both normalizations.

An unparticle mass can not be greater than the scale of the phase transition $\lambda_u$, above which the effective unparticle description is not valid.
This means that for the unit normalization, an unparticle mass can not be greater than $\Lambda_u$, since $\lambda_u = \Lambda_u$.
However, for the general normalization where $N$ can be large, from Eq.~(\ref{gn}) we see that for the fixed $d_u$, $\lambda_u \gg \Lambda_u$, so unparticle mass can be much greater than $\Lambda_u$.

\begin{figure}[t]
   \centering
\includegraphics[width=3in]{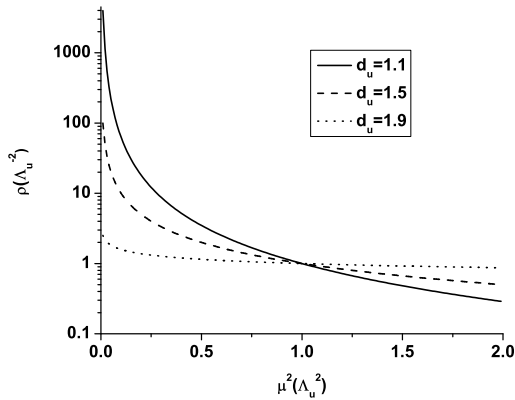}
\caption{The case of the general normalization (many unparticle degrees of freedom). The figure shows  the mass distribution $\rho(\mu)$ of unparticles for the fixed normalization scale $\Lambda_u$. For convenience, $\rho$ is measured in units of $\Lambda_u^{-2}$ while $\mu$ is measured in units of $\Lambda_u$. Since the scaling dimension $d_u$  can take the values between $1$ and $2$, we plot the three characteristic values: $d_u=1.1$ solid line, $d_u=1.5$ dashed line and $d_u=1.9$ doted line. As the value of the scaling dimension decreases, the distribution is skewed toward lighter unparticles.}
    \label{dufixed}
\end{figure}

Obviously, the mass distribution of unparticles, depends on the scaling dimension $d_u$ and the normalization scale  $\Lambda_u$ in a non-trivial way. To get some intuition for the case of the general normalization, in Fig.~\ref{dufixed} we plot the mass distribution $\rho(\mu)$ for three different values of $d_u$. For convenience, $\rho$ is measured in units of $\Lambda_u^{-2}$ while $\mu$ is measured in units of $\Lambda_u$. [Alternatively, we could make the plots while keeping $\Lambda_u$ fixed.]  Fig.~\ref{dufixed} indicates that as the value of the scaling dimension decreases, the distribution is skewed toward lighter unparticles. This means that for smaller $d_u$, there are more lighter unparticle degrees of freedom.

\begin{figure}[t]
   \centering
\includegraphics[width=3in]{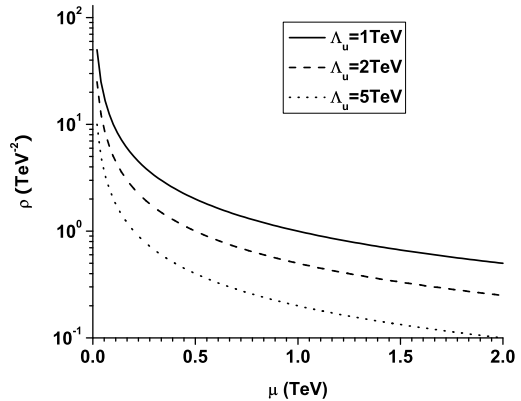}
\caption{The case of the general normalization (many unparticle degrees of freedom). The figure shows the mass distribution $\rho(\mu)$ of unparticles for the fixed $d_u=1.5$. We plot the distribution for several values of $\Lambda_u$: $\Lambda_u=1$TeV solid line, $\Lambda_u=2$TeV dashed line and $\Lambda_u=5$TeV doted line. As the value of $\Lambda_u$ decreases, there are more unparticle degrees of freedom (for a fixed $d_u$).}
    \label{Lfixed}
\end{figure}

In Fig.~\ref{Lfixed}, we also plot the mass distribution $\rho(\mu)$ for three different values of $\Lambda_u$, while we keep $d_u$ fixed, again for the case of the general normalization. We see that for smaller $\Lambda_u$  there are more unparticle degrees of freedom.
Since in this case $\lambda_u \gg \Lambda_u$, we can plot the curves up to the energy values much higher than $\Lambda_u$, which will not be the case for the unit normalization.

\begin{figure}[t]
   \centering
\includegraphics[width=3in]{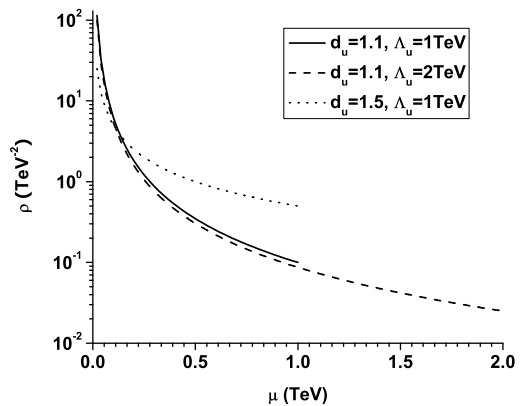}
\caption{The case of the unit normalization (a single unparticle degree of freedom). The figure shows the mass distribution $\rho(\mu)$ for the three characteristic pairs of values: $d_u=1.1$, $\Lambda_u=1$TeV solid line, $d_u=1.1$, $\Lambda_u=2$TeV dashed line and $d_u=1.5$, $\Lambda_u=1$TeV doted line. As the value of the scaling dimension decreases, the distribution is skewed toward lighter unparticles. In this case unparticles can not be heavier than $\Lambda_u$, so the distribution stops there.}
    \label{du-normal}
\end{figure}

We now study the case of the unit normalization. Fig.~\ref{du-normal} shows the mass spectra for the three characteristic pairs of values: $d_u=1.1$, $\Lambda_u=1$TeV solid line, $d_u=1.1$, $\Lambda_u=2$TeV dashed line and $d_u=1.5$, $\Lambda_u=1$TeV doted line. Qualitative situation is the same as in the case of the general normalization. As the value of the scaling dimension decreases, the distribution is skewed toward lighter unparticles. However, since in this case $\lambda_u = \Lambda_u$, we can plot the curves only up to the energy values of $\Lambda_u$.

Since unparticles have effectively continuous mass, a mass of an unparticle with energy $\omega$ takes all the values from $0$ to $\omega$. Therefore, we need to integrate over all the possibilities. Thus, the power spectrum must be written as
\begin{equation} \label{int}
\frac{d^2 E}{dtd\omega} = \int^{\omega^2}_0 \rho (\mu) \frac{d^2 E_\mu}{dt d
\omega}d\mu^2
\end{equation}
where $E_\mu$ is given by Eq.~(\ref{Emu}).
Eq.~(\ref{int}) is then integrated numerically.

\section{Results}

\subsection{$3+1$ dimensional black holes}

\begin{figure}[t]
   \centering
\includegraphics[width=3in]{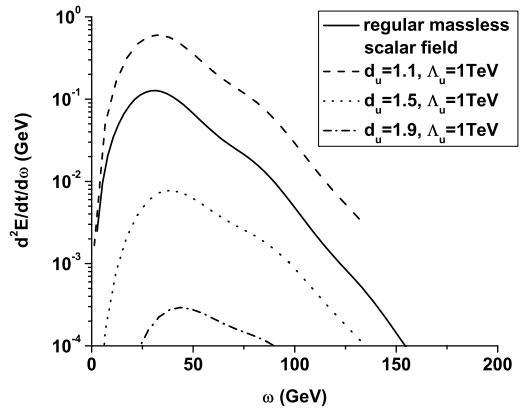}
\caption{The Hawking radiation power spectra of unparticles with the fixed value of $\Lambda_u$ (general normalization). The mass of the black hole is $10^{12}$g and the unparticle energy scale is $\Lambda_u =1$TeV. Since the scaling dimension $d_u$  can take the values between $1$ and $2$, we plot the three characteristic values: $d_u=1.1$ dashed line, $d_u=1.5$ doted line and $d_u=1.9$ dash-dot line. For comparison, we plot the regular scalar field with the solid line. Scalar unparticle spectrum can clearly be different from the ordinary scalar particle spectrum.}
    \label{spectra}
\end{figure}

In Fig.~\ref{spectra} we plot the power spectrum obtained by integrating Eq.~(\ref{int}) for the case of the general normalization given in Eq.~(\ref{gn}). The particular value of the black hole mass is chosen to be $10^{12}$g and the scale $\Lambda_u =1$TeV. Since the scaling dimension $d_u$  can take the values between $1$ and $2$, in order to cover the whole parameter space we plot the three characteristic values: $d_u=1.1$ with the solid line, $d_u=1.5$ with the dashed line and $d_u=1.9$ with the doted line. For comparison, we also plotted the power spectrum of the regular scalar field with the dash-dot line. Obviously, scalar unparticle spectrum can clearly be different from the ordinary scalar particle spectrum, and unparticles can easily out-power ordinary particles. The power emitted in unparticles grows as the scaling dimensions gets closer to $d_u=1$, and at a certain point exceeds the power emitted in ordinary scalar particle. For these values, the black hole Hawking radiation is dominated by unparticles.
For example, the emitted unparticle power for $d_u=1.1$ is about $10$ times greater than power emitted in ordinary scalar field. This effect may significantly shorten the expected life-time of a black hole.

The fact that unparticle power grows as the scaling dimension goes down may be explained by observing that lower scaling dimensions imply lighter unparticles, which are in turn easier emitted by a black hole at a given temperature.

From Fig.~\ref{spectra} we also see that most of the power is emitted at frequencies close to $\omega \sim r_g^{-1}$, where $r_g$ is the gravitational radius of the black hole. [For the chosen black hole of $10^{12}$g, we have $r_g^{-1} \sim 50$GeV.] This feature is expected since the peak of emitted power is determined by the temperature of the black hole which roughly goes as $r_g^{-1}$.


\begin{figure}[t]
   \centering
\includegraphics[width=3in]{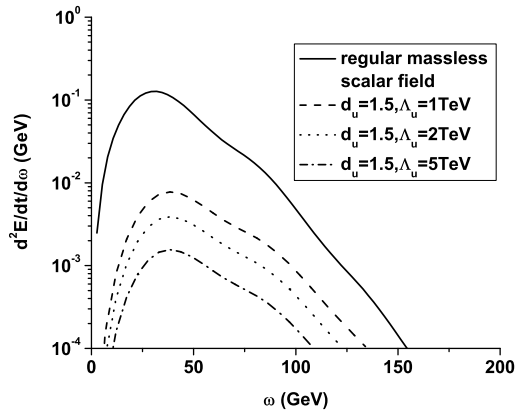}
\caption{The Hawking radiation power spectra of unparticles for several values of $\Lambda_u$ and fixed $d_u=1.5$ (general normalization). The mass of the black hole is $10^{12}$g. We plot $\Lambda_u=1$TeV dashed line, $\Lambda_u =2$TeV doted line and $\Lambda_u=5$TeV dash-dot line. The emitted power in unparticles is larger for smaller $\Lambda_u$ since then we have more unparticle degrees of freedom at a given energy. For comparison, we plot the regular scalar field with the solid line. Again, scalar unparticle spectrum can clearly be different from the ordinary scalar particle spectrum.}
    \label{lambda}
\end{figure}

\begin{figure}[t]
   \centering
\includegraphics[width=3in]{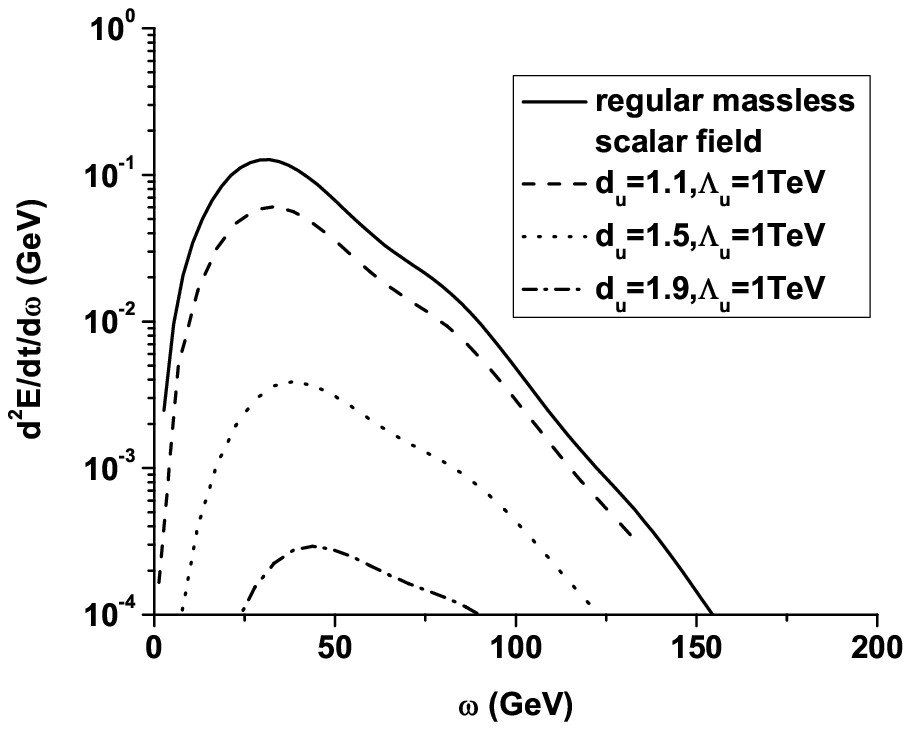}
\caption{The Hawking radiation power spectra of unparticles with the fixed $\Lambda_u$ (unit normalization). The mass of the black hole is $10^{12}$g and the unparticle energy scale is $\Lambda_u =1$TeV. Since the scaling dimension $d_u$  can take the values between $1$ and $2$, we plot the three characteristic values: $d_u=1.1$ dashed line, $d_u=1.5$ doted line and $d_u=1.9$ dash-dot line. For comparison, we plot the regular scalar field with the dash-dot line. Scalar unparticle spectrum can clearly be different from the ordinary scalar particle spectrum. As $d_u$ is approaching $1$, the unparticle spectrum approaches that of a single massless scalar field.}
    \label{normalize}
\end{figure}

\begin{figure}[t]
   \centering
\includegraphics[width=3in]{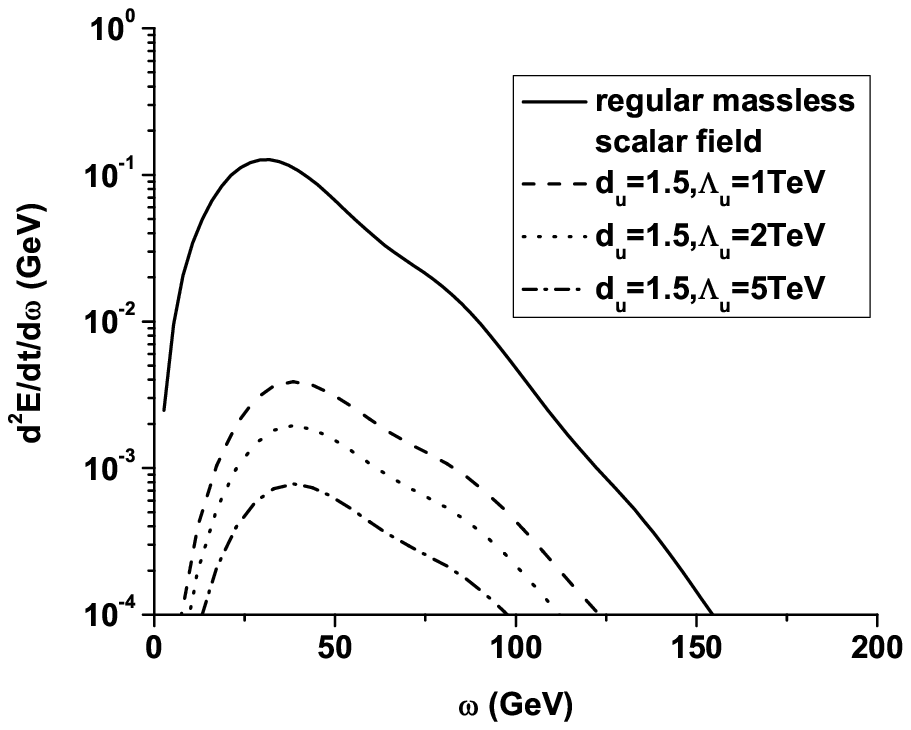}
\caption{The Hawking radiation power spectra of unparticles for several values of $\Lambda_u$ and fixed $d_u=1.5$ (unit normalization). The mass of the black hole is $10^{12}$g. We plot $\Lambda_u=1$TeV dashed line, $\Lambda_u =2$TeV doted line and $\Lambda_u=5$TeV dash-dot line. The emitted power in unparticles is larger for smaller $\Lambda_u$ since the mass distribution $\rho(\mu)$ gives higher probability for a lighter unparticle. For comparison, we plot the regular massless scalar field with the solid line. Again, scalar unparticle spectrum can clearly be different from the ordinary scalar particle spectrum.}
    \label{normalize-1}
\end{figure}

In Fig.~\ref{lambda} we plot the Hawking radiation power spectra of unparticles for several values of $\Lambda_u$ and fixed $d_u=1.5$, for the case of the general normalization given in Eq.~(\ref{gn}). As the value of $\Lambda_u$ goes down, the emitted power in unparticles goes up. The reason is that for smaller $\Lambda_u$, there are more unparticle degrees of freedom at a given energy (i.e temperature of the black hole), as can be seen from Fig.~\ref{Lfixed}. For comparison, we plot the regular scalar field with the dash-dot line. Again, scalar unparticle spectrum can clearly be different from the ordinary scalar particle spectrum.

It is worth noting that, in the limit if $d_u \rightarrow 1$, we do not recover the standard scalar field result, which is a direct consequence of the form of the general normalization in Eq.~(\ref{gn}) which implies that in the limit of $d_u \rightarrow 1$ we have many scalar unparticle degrees of freedom. However, with the choice of the unit normalization in Eq.~(\ref{rhou}), $\rho$ does reduce to the Dirac delta function in $d_u \rightarrow 1$ limit and the total number of unparticle degrees of freedom becomes $1$ (a single massless scalar field). This implies that in $d_u \rightarrow 1$ limit of the unit normalization we should recover the Hawking radiation spectrum for an ordinary massless scalar field.

In Fig.~\ref{normalize} we plot the power spectrum obtained by integrating Eq.~(\ref{int}) for unparticles with the unit normalization given in Eq.~(\ref{rhou}). The particular value of the black hole mass is chosen to be the same as in Fig.~\ref{spectra}. Here we make plots with the fixed $\Lambda_u$ and several values of $d_u$. As indicated in the previous paragraph, as $d_u$ is getting closer to $1$, the spectra become closer to those for the standard massless scalar field. With this normalization, unparticles can not out-power ordinary particles.

In Fig.~\ref{normalize-1} we plot the power spectrum for unparticles with the unit normalization, with the fixed $d_u =1.5$ and several values of $\Lambda_u$. The particular value of the black hole mass is chosen to be the same as for the other plots. The qualitative features are similar as in the general normalization.  The emitted power in unparticles is larger for smaller $\Lambda_u$ since  the mass distribution $\rho(\mu)$ gives higher probability for a lighter unparticle.

\subsection{Higher dimensional black holes}

Small black holes whose Hawking radiation may in principle be tested are of great interest in theories with TeV scale gravity with extra dimensions. With an inception of the Large Hadron Collider (LHC) it is very important to study eventual experimental signature of mini black hole production and their subsequent evaporation. A lot of work has been done on this topic \cite{Kanti:2004nr,Landsberg:2006mm,Stojkovic:2004hp,Dai:2006hf,Stojkovic:2005zq}, including building  very comprehensive black hole event generators \cite{Dai:2007ki}, however, no unparticle signature has been studied so far.

\begin{figure}[t]
   \centering
\includegraphics[width=3in]{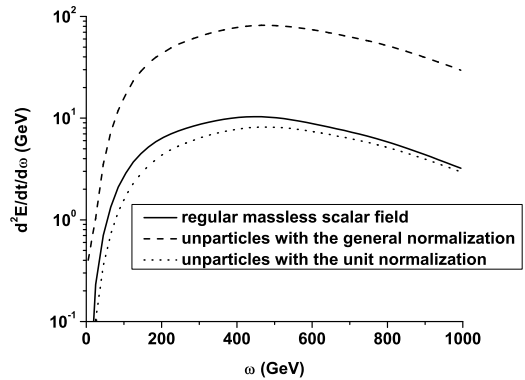}
\caption{The Hawking radiation power spectrum of unparticles for a higher dimensional black hole. The black hole mass is $5$TeV, the number of extra dimensions is $n=3$, while the higher dimensional Planck mass is $1$TeV. The unparticle energy scale is $\Lambda_u =1$TeV and we take $d_u=1.1$. The unparticle spectra are represented by the dashed line and dotted line. For comparison we plot the regular massless scalar field with the dashed line. Clearly, unparticles with the general normalization can easily out-power ordinary particles by an order of magnitude. Unparticles with the unit normalization can only asymptotically approach the ordinary scalar field in $d_u \rightarrow 1$ limit.}
    \label{spectra-high}
\end{figure}

Calculation of gray-body factors for the higher dimensional black holes closely follows the method we developed for $3+1$ dimensions.  The only change is in the metric (\ref{metric}). The term $F(r)$ is now
\be
F(r)=1-\left( \frac{r^{(n)}_g}{r} \right)^{1+n}
\ee
where $r^{(n)}_g$ is higher dimensional gravitational radius, while $n$ is the number of extra dimensions.
In Fig.~\ref{spectra-high} we plot the power spectrum obtained by integrating Eq.~(\ref{int}). The particular value of the black hole mass is chosen to be $5$TeV, the number of extra dimensions is $n=3$, while the higher dimensional Planck mass is $1$TeV. We also set $\Lambda_u =1$TeV and $d_u=1.1$.

It is natural to expect that interesting effects happen if  $\Lambda_u$ (and therefore $\lambda_u$) is not much different than higher dimensional Planck scale. Indeed, for the set of parameters we chose here, the power emitted in scalar unparticles is about $10$ times larger than the power emitted in the ordinary massless scalar field. Qualitative features remain the same as in $3+1$ dimensional case, i.e. the unparticle effects become more prominent as $\Lambda_u$ is getting smaller. Unparticle effects are also enhanced as $d_u$ is getting closer to $1$ in the case of the general normalization since there are many unparticle degrees of freedom with that normalization. For the unit normalization, the unparticle spectrum approaches the standard one as $d_u$ is getting closer to $1$.
Thus, unparticle sector can significantly modify experimental signature of mini black holes production at the LHC. Because of the very weak interactions between the standard model particles and unparticles, most of the emitted energy might go into the hidden sector. This implies a large percentage of missing energy in high energy collisions, especially in the case of the general unparticle normalization.

\section{Conclusions}

We studied the consequences of an eventual existence of unparticle degrees of freedom for the black hole Hawking radiation. Unparticle sector is characterized with several quantities. The fundamental energy scale of the new scale invariant sector is $M_F$. This scale is supposedly high above energies accessible in colliders. However, its effects may be observed as unparticles. The scale of the phase transition above which the effective unparticle description is not valid is $\lambda_u$, the characteristic (normalization) scale of unparticle degrees of freedom is $\Lambda_u$ (with the hierarchy $\Lambda_u \leq \lambda_u < M_F$). The scaling dimension of unparticles is $d_u$ with the property $1 < d_u \leq 2$. Another peculiar property of the unparticle degrees of freedom is that their mass is continuous.

We developed a method for calculating the gray-body factors for the scalar unparticle mode emitted by $3+1$ dimensional and higher dimensional black holes.
Our results shows that the unparticle spectra are very different from the spectra of ordinary particles. In general, the power emitted in unparticles grows as the value of the scaling dimension $d_u$ gets closer to $1$ (in the case of the general normalization). The power is also larger for smaller values of  $\Lambda_u$, regardless of the particular normalization. In the first case the mass distribution of unparticles is shifted toward smaller values, which implies that there are more lighter unparticle degrees of freedom. Similarly, in the second case, the mass distribution $\rho(\mu)$ gives higher probability for a lighter unparticle.  A black hole at a given temperature emits all of the degrees of freedom which are lighter than its temperature, which explains the $d_u$ and $\Lambda_u$ dependence of the Hawking radiation.


We compared the power emitted in unparticles with the power emitted in ordinary scalar field particles and found that the unparticle power may be larger by many orders of magnitude for both  $3+1$ dimensional and higher dimensional black holes. This may significantly affect the black hole life-time, which in turn can modify the phenomenology of primordial black holes. Unfortunately, it is difficult to obtain an exact quantitative result without studying the grey-body factors for all of the unparticle degrees of freedom (e.g. vector, tensor etc.).

In the context of TeV scale gravity models with extra dimensions, a dominant unparticle emission may significantly modify the collider signature of mini black hole production.
In particular, unparticles imply a large percentage of missing energy in high energy collisions.

Finally, we pointed out that some results are sensitive to the choice of the normalization factors for unparticles. At this point, exact combinatoric factors for unparticles are unknown, and the choice of normalization is purely conventional.

\begin{acknowledgments} DS acknowledges the financial support from NSF, award number PHY-0914893.
\end{acknowledgments}

\end{document}